\documentclass[12pt]{article}
\usepackage{epsfig,graphics,amssymb}
\newcommand{\bear}{\begin{equation}\begin{array}}
\newcommand{\eear}[1]{\end{array}\label{#1}\end{equation}}
\newcommand{\be}{\begin{equation}}
\newcommand{\ee}{\end{equation}}
\newcommand{\bes}{\begin{subequations}}
\newcommand{\ees}{\end{subequations}}
\newcommand{\bea}{\begin{eqnarray}}
\newcommand{\eea}{\end{eqnarray}}
\newcommand{\nn}{\nonumber}
\def\ba{$$\begin{array}}
\def\ea{\end{array}$$}
\def\bra{$\begin{array}}
 \def\era{\end{array}$}

\def\Re{\mathop{\mathrm{Re}}\nolimits}
\def\Im{\mathop{\mathrm{Im}}\nolimits}

\def\ra{\rightarrow}
\def\to{\rightarrow}

\def\half{\frac{1}{2}}

\begin{document}
\setlength{\textwidth}{16.cm}
\setlength{\textheight}{23cm}

\oddsidemargin 1cm
\evensidemargin -0.2cm
\addtolength{\topmargin}{-2.5 cm}

\begin{flushright}
hep-ph/0512371 \\
December 2005 \\
\end{flushright}
\vskip 1cm

\centerline{\Large{Testing Higgs sector of 2HDM}}

\vskip 1cm
\centerline{\large Maria Krawczyk \footnote{
{ Presented at International Europhysics Conference on High Energy Physics,
                 July 21st - 27th 2005,
                 Lisboa, Portugal}}}
\vskip 0.5cm

 \centerline{Institute of Theoretical Physics, Warsaw University, Poland}


\abstract
{Properties of the Higgs sector of
Two Higgs Doublet Model (2HDM)
and existing constraints on its parameters are discussed. Potential of 
the  Photon Linear Collider in testing various Higgs scenarios of 2HDM,
including the MSSM, based on the realistic simulations is also presented.
}


\section{Symmetries of the 2HDM \cite{Ginzburg:2004vp}}
The Higgs sector with two scalar complex doublets ($Y=\pm 1$)
is the simplest yet very rich extension of the Standard Model (SM). 
The general 2HDM potential is 
\bea
&V=\half\lambda_1(\phi_1^\dagger\phi_1)^2
+\half\lambda_2(\phi_2^\dagger\phi_2)^2 
+\lambda_3(\phi_1^\dagger\phi_1) (\phi_2^\dagger\phi_2)\\ \nonumber
&+\lambda_4(\phi_1^\dagger\phi_2) (\phi_2^\dagger\phi_1) 
+\half\left[{\lambda_5}(\phi_1^\dagger\phi_2)^2+{\rm h.c.}\right]\\ \nonumber
&+\left\{\left[{\lambda_6}(\phi_1^\dagger\phi_1)
+{\lambda_7}(\phi_2^\dagger\phi_2)\right](\phi_1^\dagger\phi_2)
+{\rm h.c.}\right\} \\ 
&-\half \bigl\{m_{11}^2(\phi_1^\dagger\phi_1)
+{\left[m_{12}^2 (\phi_1^\dagger\phi_2)+{\rm h.c.}\right]}
+m_{22}^2(\phi_2^\dagger\phi_2)\bigr\}, \nonumber
\eea
with possible complex coefficients:
{$\lambda_5, \lambda_6, \lambda_7$}, and $m_{12}^2$ (i.e. in total  
14 real parameters). 
{{There is no $(\phi_1,\phi_2)$ transition  {{(as a consequence there is no 
FCNC nor CPV)}} if $V$ is 
{{$Z_2$ symmetric}} under transformations:  
$\phi_1\to -\phi_1$, $\phi_2\to \phi_2$ (or vice versa). This means that 
$\lambda_6=\lambda_7=m_{12}^2=0$}}.
Hard violation of $Z_2$ symmetry typically means appearance of
 quartic terms with {$\lambda_6$, $\lambda_7$}, while
 soft violation of $Z_2$ symmetry ({$\lambda_6,\lambda_7=0$}) is 
 governed by  {{$  m_{12}^2$}}.

Two fields with identical quantum numbers can mix without changing a 
physical picture. Such mixing is described by
a global unitary transformation U(1) x SU(2):
\bea
\left( \begin{array} {c}  \phi_1'\\
{ \phi}_2' \end{array}\right) = e^{-i\rho_0} \left( \begin{array} {cc}
{{\cos\theta}} \,e^{i{{\rho}}/2}&{{\sin\theta}}\,e^{i(\tau-\rho/2)}\\
-\sin\theta\,e^{-i(\tau-\rho/2)}&\cos\theta\,e^{-i\rho/2}
\end{array}\right) \left( \begin{array}{c} \phi_1\\
\phi_2\end{array}\right) \, .
\eea
It induces changes in  the parameters  of potential $V$ and the whole 
Lagrangian $L$.

{{Reparametrization transformations (RPaT)}: $\lambda_i\to\lambda'_i$ and  
$m_{ij}^2\to (m')_{ij}^2$},  constitute a 3 parametrical  group with  
parameters:  {$\rho$, $\theta$, $\tau$} (here $\rho_0$ is not relevant)
acting in the 14-dim space of Lagrangians. 
(The rephasing transformation group ($\theta=\tau=0$) 
has one parameter only {- $\rho$}.)
As various Lagrangians are  reparametrization invariant
there is a reparametrization equivalent 3-dim subspace of Lagrangians.
{Of course physical observables, like masses,
 are invariant under  the RPaT.} 
A CP violation  in the Higgs sector is signalized by the 
complex parameters of the Lagrangian.

The U(1)$_{QED}$ symmetric vacuum 
 corresponds to the lowest energy than the ``charged vacuum''; 
it can be choosen as
\be
\langle\phi_1\rangle =\frac{1}{\sqrt{2}}\left(\begin{array}{c} 0\\
v_1\end{array}\right), \;\; \langle\phi_2\rangle
=\frac{1}{\sqrt{2}}\left(\begin{array}{c}0 \\[5mm] v_2
 e^{i\xi}\end{array}\right).\label{genvac}\ee
The rephasing of fields  can always remove  the phase
difference  $\xi\to\xi-\rho\,$. 
Instead of the oryginal 
parameters $m_{ij}^2$ from eq.(1) one can introduce $v_1, v_2$ 
(with $v_1^2+v_2^2=v^2, v$ = 246 GeV) and  $ \nu=\Re m_{12}^2/2v_1v_2$.
Note, that the famous  parameter {{$\tan \beta={v_2}/{v_1}$} }  
depends on reparametrization!

{{In 2HDM there are 5 physical Higgs bosons: three  neutral ones
$h_1,h_2,h_3$  and two charged $H^\pm$}}.
If there is no CP violation physical neutral particles $h,H,A$ have definite 
CP properties,  otherwise  $h_1,h_2,h_3$ are mixtures of  $h,H,A$ states.
Small or large mixing angles lead, respectively, to weak or strong CP violation
(by mixing). 

The Yukawa Lagrangian  for quarks (similarly fo leptons) is
  \bear{c}
-{\cal L}_{\rm Y} =\bar Q_L [(\Gamma_1\phi_1+\Gamma_2\phi_2)
d_R
+(\Delta_1\tilde\phi_1+\Delta_2\tilde\phi_2) u_R] +{\rm h.c.}.
 \end{array}\nn
 \ee
$\Gamma_1=\Delta_2=0$ with diagonal real $\Gamma_2, \Delta_1 $ 
corresponds to so called  Model II (MSSM is of this type).  

 We argue that   2HDM (II) with  {\it weak} CP violation ({\it small} mixing 
angles), soft $Z_2$ violation and small  $\nu$  is natural. This means that  
non-decoupling  of heavy Higgs bosons, absent for large $\nu$,  
is a natural property of 2HDM, in contrast to MSSM.

There are various relations between physical couplings of Higgs bosons
to other particles, which can be used to
test the model.  It is useful to introduce relative couplings 
{${\chi_j^{(i)}= {g_j^{(i)}}/{g_j^{\rm SM}}}$} ($\,\, j=V,u,d$),
with  sum rules
 $\Sigma_i (\chi_j^{(i)})^2 = 1.$ 
{ In addition there is a pattern relation, which   hold
for  {\it each neutral Higgs particle $h_i$}
(in particular also for $h,H,A$ in the case of CP conservation):
 \bear{c}\label{2hdmrel}
(\chi_u^{(i)} +\chi_d^{(i)})\chi_V^{(i)}=1+\chi_u^{(i)}
\chi_d^{(i)}\,\,\,,
(\chi_u^{(i)}-\chi_V^{(i)})
(\chi_V^{(i)}-\chi_d^{(i)})=1-(\chi_V^{(i)})^2.\end{array}
\end{equation}
One can determine {{$\tan \beta=v_2/v_1=\tan \beta_{II}$}}
 via equations which hold in 2HDM (II) for $h_i$ and also for
$h,H,A$ (except the last one, absent for $h,H$):
\begin{equation}
\!\!\tan^2\beta_{II}\!=
\!{\frac{(\chi_V^{(i)}\!-\!\chi_d^{(i)})^*}{\chi_u^{(i)}\!-\!\chi_V^{(i)}}}
\!=\! {\frac{1\!-\!|\chi_d^{(i)}|^2}{|\chi_u^{(i)}|^2\!-\!1}}= \!
\frac{\Im \chi_d^{(i)}}{\Im \chi_u^{(i)}}\!. \nn
\end{equation}
Neutral Higgs bosons couple to photons  via loops with 
all charged particles existing in the theory. For the effective vertex
$h_i\gamma \gamma$ 
the coupling $h_i H^+ H^-$ contributes: 
\bear{c}
\chi_{H^\pm}^{(i)}
=\left(1-\frac{M_i^2}{2M_{H^\pm}^2}\right)\chi_V^{(i)}
+\frac{M_i^2-\nu v^2}{2M_{H^\pm}^2}
\Re(\chi_u^{(i)}+\chi_d^{(i)}).
 \end{array}
\end{equation}
For {{small $\nu$}} and large $M_{H^\pm}$ even if  all  $\chi_V,\chi_u,\chi_d$ 
are equal 1
 { {(SM-like scenario)}} $\chi_{H^\pm}^{(i)} \approx 1$  there may occur
{{large non-decoupling}} effects due to $H^\pm$. Contrary, for 
{large $\nu \sim M_{H^\pm}^2/v^2$ one gets $\chi_{H^\pm}^{(i)} \approx 0$. }

\section{Constrainig 2HDM from the leptonic tau decays \cite{Krawczyk:2004na}}
The CP conserving 2HDM was tested at LEP in:
Bjorken process $Z\rightarrow Zh$  (constraining $\chi_V^{(h)}=\sin(\beta-\alpha)$),
pair production  $e^+e^-\rightarrow hA$ ($\cos(\beta-\alpha)$),
Yukawa process $e^+e^-\rightarrow bbh/A$, $\tau\tau h/A$ ($\chi_d^{(h/A)}$). 
Also
$e^+e^-\rightarrow H^+H^-$ and  a loop process  $Z^*\rightarrow h/A \gamma$  
were used to constrain various couplings and masses. 
Still, one very light neutral Higgs boson $h$ (with small $\chi_V$)
or $A$ may exist with relatively large Yukawa coupling to $b, \tau$.
Also precise  $(g-2)_\mu$ data 
{do not rule out such  light Higgs scenario in 2HDM
 (e.g. for small $\nu $ and 
light $A$, with $\chi_d \in $ between 40-100).} 
Note, that ${H^{\pm}}$ has to be heavier than  490 GeV,
as follows from  the analysis of the process  $b\rightarrow s \gamma$ 
\cite{Gambino:2001ew}.

New constraints of 2HDM  were obtained  from 
the leptonic tau decays using   the '04 world data. 
In SM these decays
proceed at  the tree-level via $W^\pm$ exchange.  
In 2HDM   {{charged-Higgs-boson} can be  exchanged} at  tree-level,
in addition.
We found that 2HDM loop corrections, 
involving also {{neutral Higgs bosons}}, 
dominate at large $\tan \beta$, giving contribution to the branching ratios
proportional to
$\sim  \tan \beta ^2\,\,[\ln {M_{h}}/{M_{H+}} + 1]$.

For  the beyond  SM contributions, defined as   
$Br^l = Br^l |_{SM} (1 + {{ \Delta^l}})$
 [$Br^l|_{SM}
=\Gamma^l|_{SM}
\tau_{\tau}^{exp} \,\,\,$],
we derived the $95\% $ C.L.  bounds for the electron and
muon decay modes:
\be{
(-0.80 \leq \Delta^e \leq 1.21) \% \,\,\,\,\,\,\,\,
(-0.76 \leq \Delta^{\mu} \leq 1.27)\% }.\ee
\begin{figure}
\includegraphics[width=.42\textwidth]{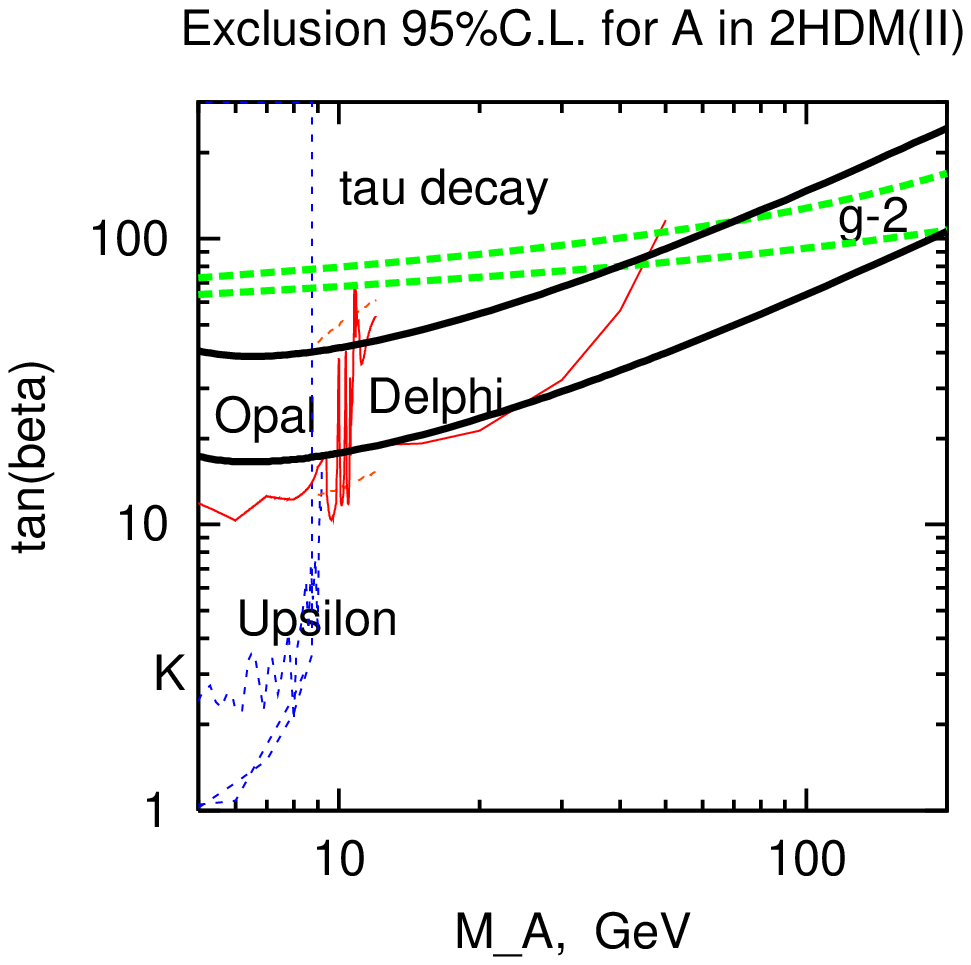}
\includegraphics[width=.42\textwidth]{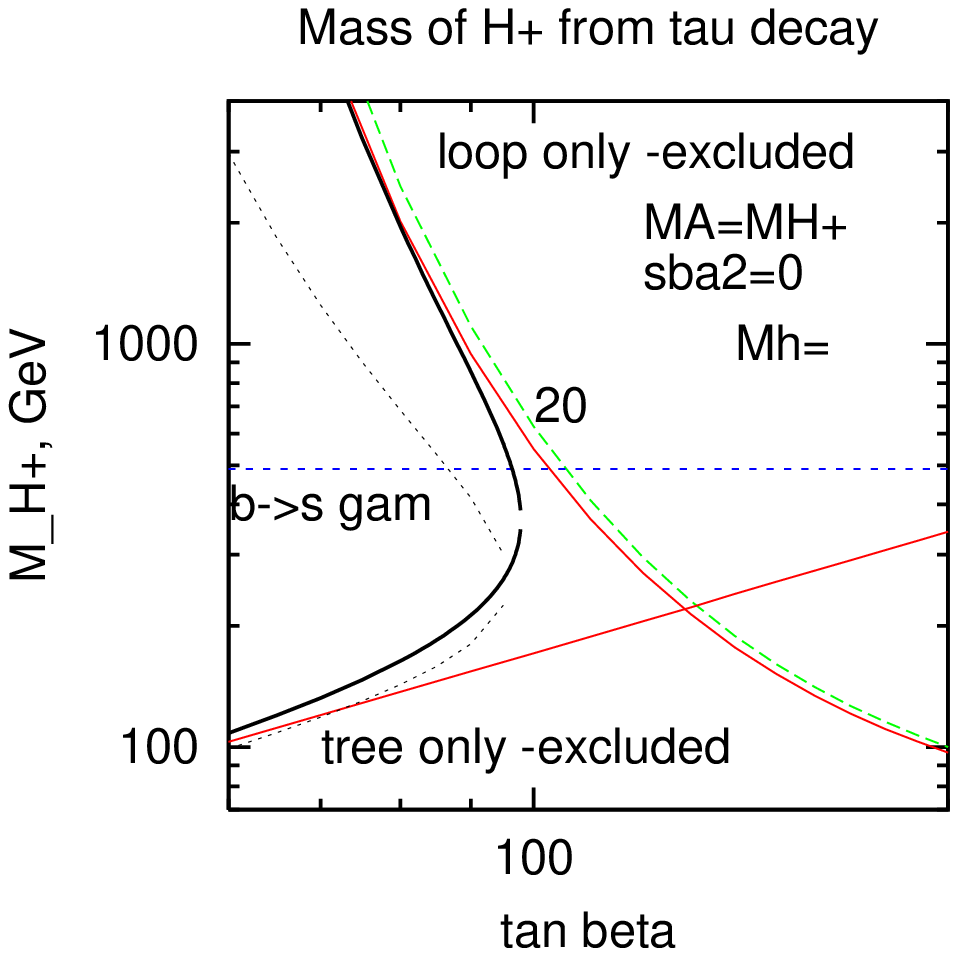}
\caption{left: Constraints for $A$; right:
 Limits $M_{H^\pm}$ versus $\tan \beta $,  
dotted line for $M_A$=100 GeV.}
\label{fig1}
\end{figure}
 From this we got {upper limits on Yukawa couplings} for both light 
$h$ and light $A$
scenarios (Fig.1,left). {New  lower  limit on mass of $M_{H^\pm}$, 
which differs significantly from  a standard constraint 
(based on the tree-level $H^{\pm}$ exchange only)} was obtained 
(Fig. 1, right). 
{We obtained also a upper limit on $M_{H^\pm}$}: 
e.g. for $\tan \beta$=80, $M_A$=100 and $M_h$=20 GeV 
mass of ${H^\pm}$ should be  below 700 GeV.
\section{Higgs-bosons production at PLC}
Higgs boson can be produced as a resonance  
at the Photon Linear Collider (PLC) in the collision of two photon beams 
obtained from the Compton backscattering processes. PLC is an ideal machine
for testing properties of neutral Higgs bosons, especially CP parities. 
Energy and polarization spectrum of photons  has to be incorporated 
in the realistic simulations.

The production relies on
loop coupling $h_i \gamma \gamma $, mentioned above. For light Higgs boson
the main decay channel is $b \bar b$, for heavier Higgses - the WW/ZZ.
In MSSM  the lightest Higgs boson has a property of that of SM, 
then, according to  the sum rules, $H/A$ decay predominately  
to $b$ quarks.

\subsection{Covering the LHC wedge at PLC \cite{Niezurawski:2005cr}}
The  state-of-art analysis of $h \ra  b \bar b$ and $A, H \ra  b\bar b$ 
bases on realistic photon energy spectra (TESLA-like) and includes effects 
of the beams crossing angle and  primary vertex distribution, as well as
the realistic detector simulation (SIMDET). 
The NLO QCD background   $Q \bar Q(g) (Q=c,b)$ as well as
backgrounds due to  $ W^+W^-, q \bar q (q=u,d, s)$  are included.
Overlaying events (OE; about 1 - 2 per bunch crossing) were taken into acount. 
Realistic b-tagging  and  corrections  for escaping  neutrinos  were applied.
\begin{figure}
\includegraphics[width=.5\textwidth]{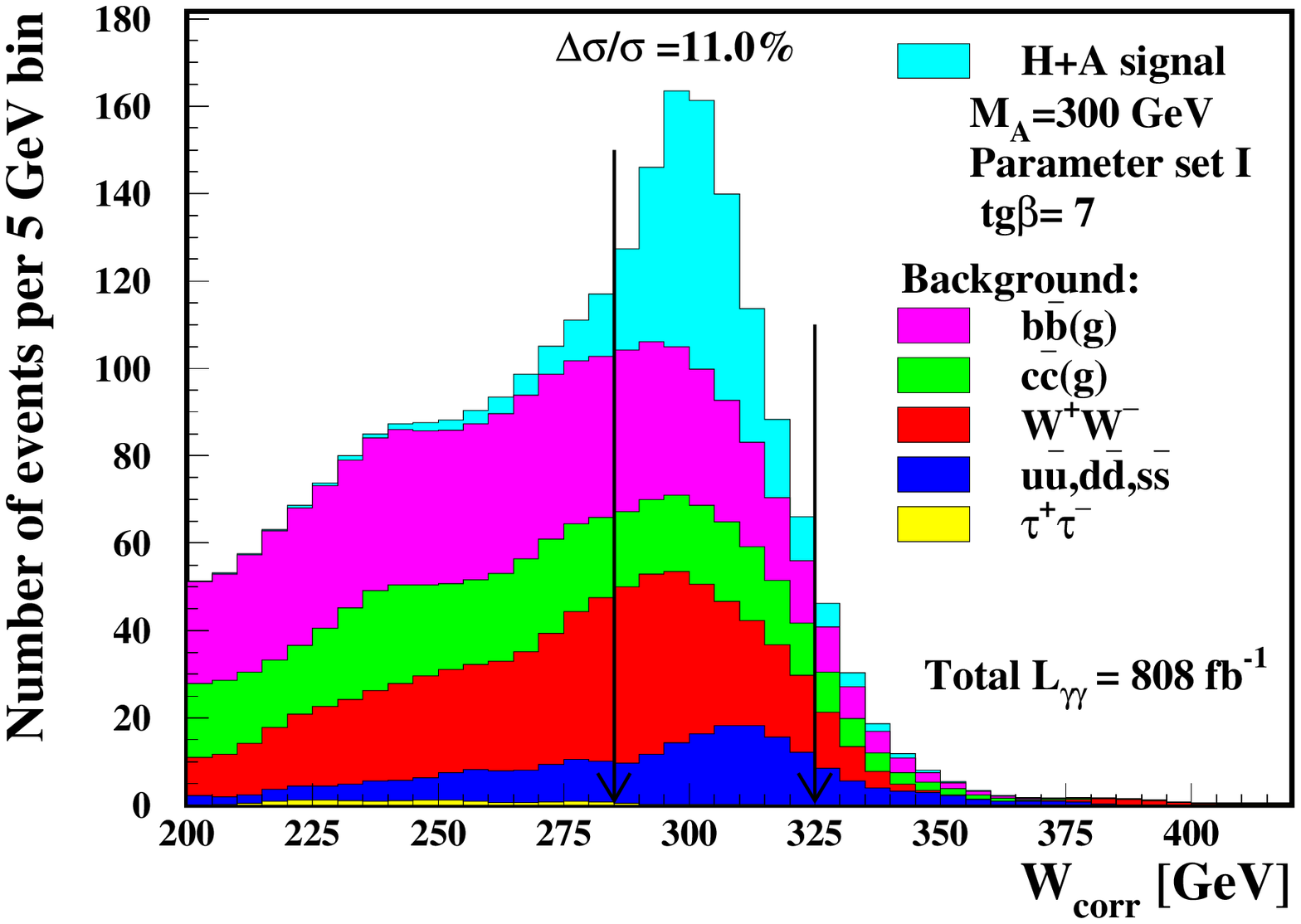}
\includegraphics[width=.4\textwidth]{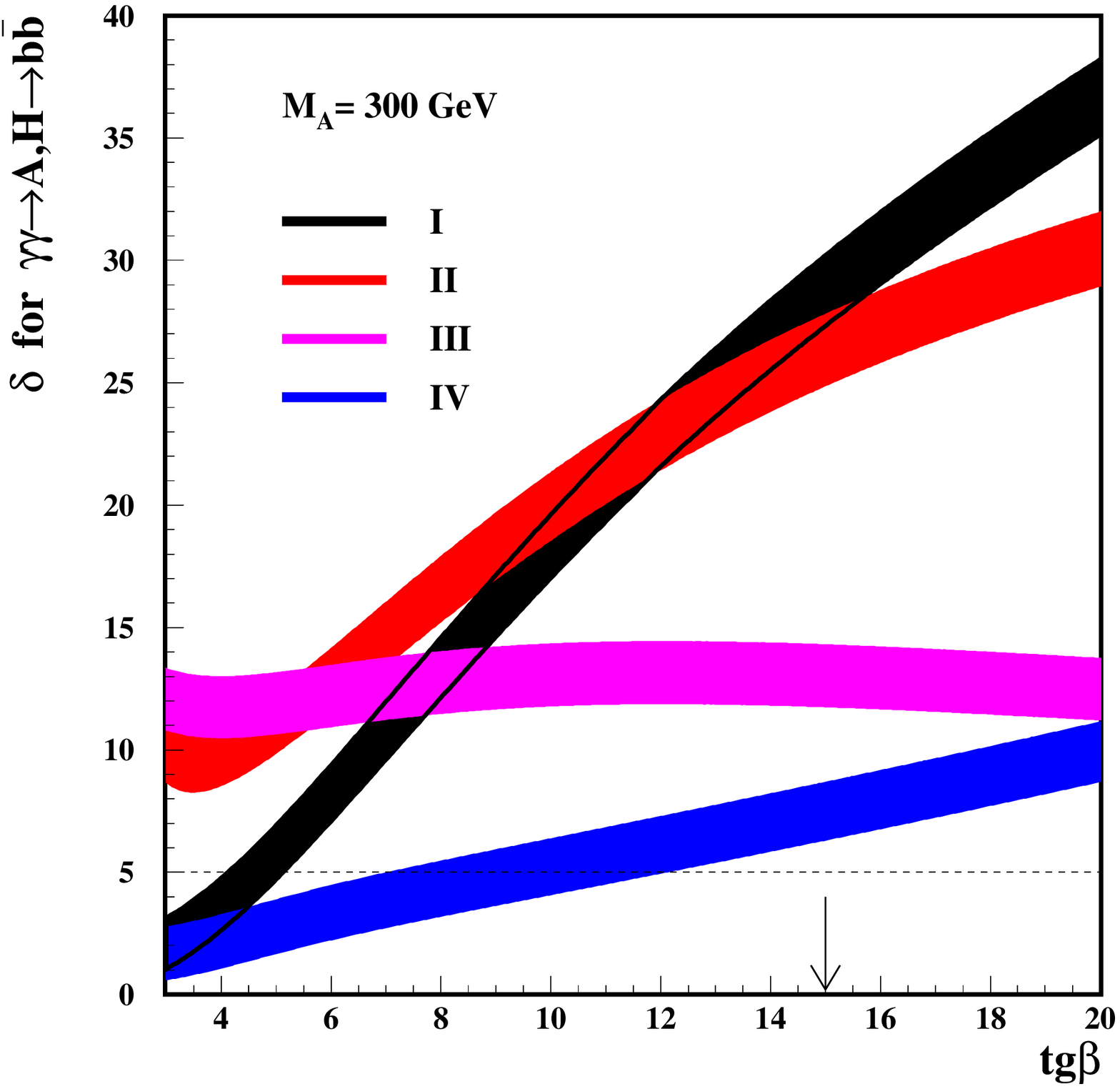}
\caption{left: Corrected inv. mass for $M_A$=300 GeV;right:Statistical 
significance as a function of $\tan \beta$.}
\label{fig2}
\end{figure}

 For the SM Higgs boson 
one obtains accuracy of the cross section measurements  of the order 2 -8\%
for $M_h$ = 120 - 160 GeV (with additional cuts for the $WW$ background).

The study for MSSM  was performed to check potential of the PLC to cover the LHC wedge, where only one SM-like $h$ can be seen. 
Four MSSM parameter sets, for $\tan \beta$ = 3 - 20
and  $M_A$ = 200 - 350 GeV  were considered.   Results are  shown in Fig. 2.
(E.g. for set I  precision is 11 - 21\% for $tan \beta$ = 7  after one year).
For $M_A$=300 GeV one  can discover MSSM Higgs bosons
below $\tan \beta$  = 15 (LHC limit) at PLC,  and therefore 
there is a chance to cover LHC wedge.
\subsection{Determination of 2HDM couplings at PLC. Combined 
 LHC, ILC and PLC analysis \cite{Niezurawski:2004ui}}
\begin{figure}
\includegraphics[width=.5\textwidth]{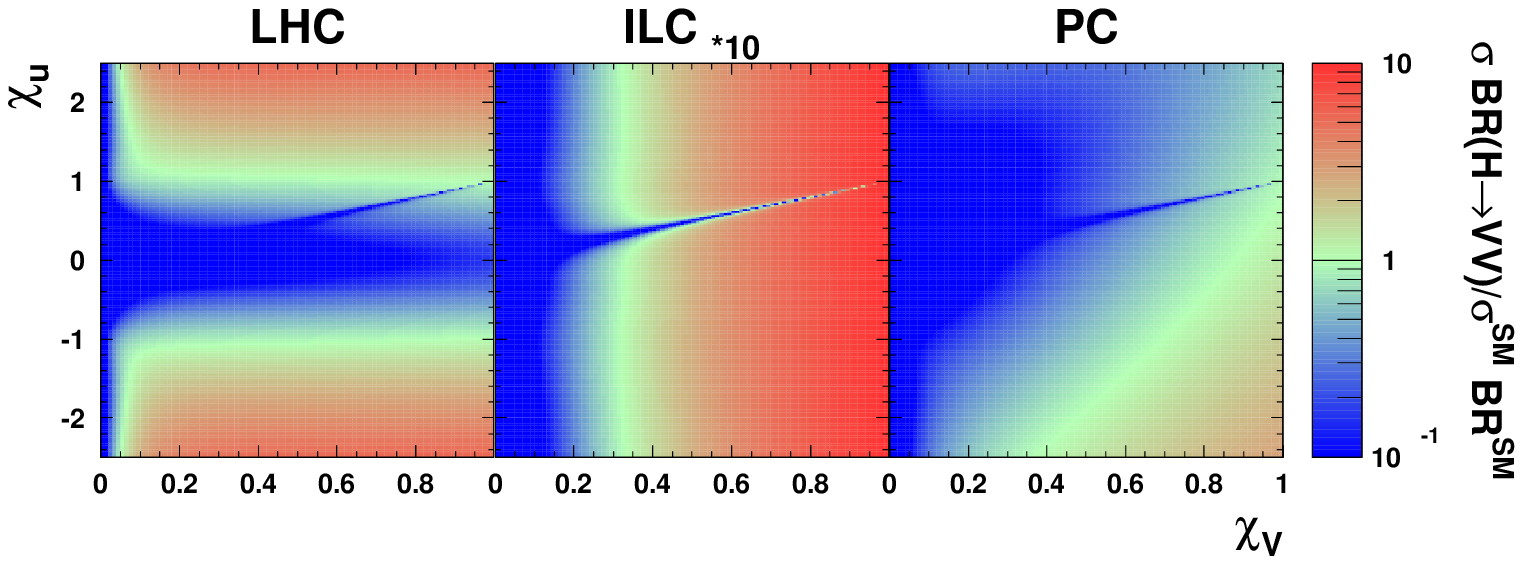}
\includegraphics[width=.5\textwidth]{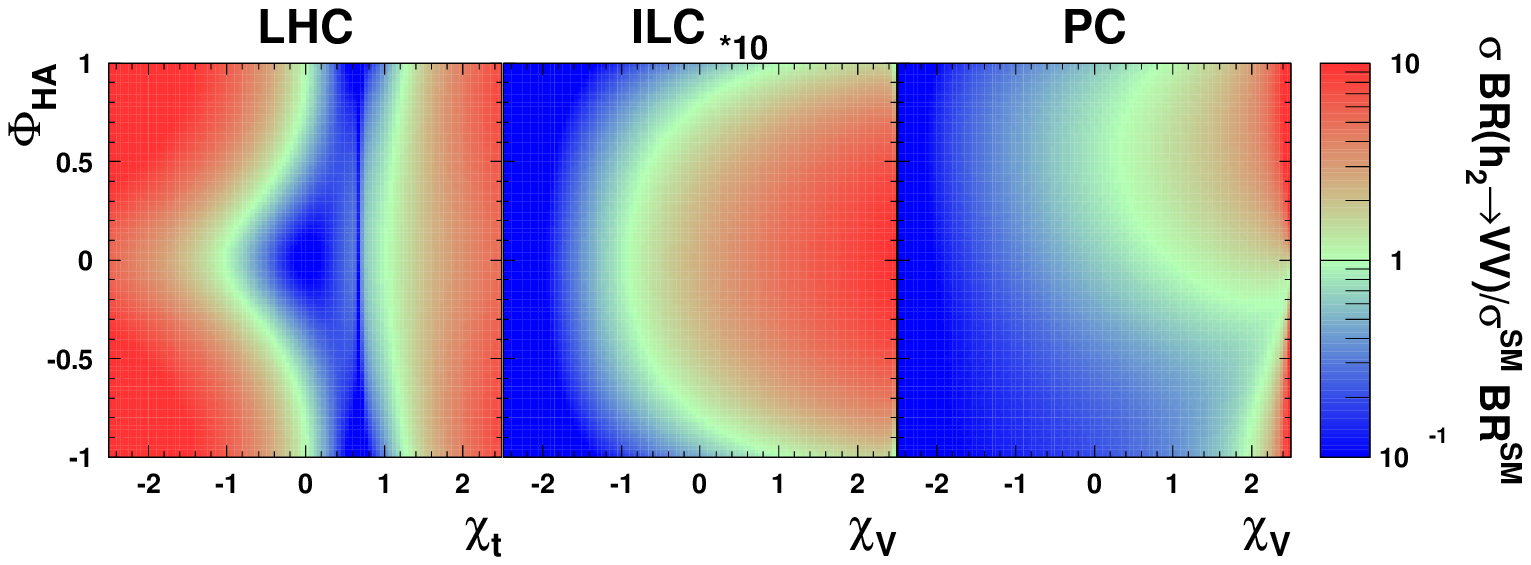}
\caption{Expected production rates for CP conserving (violating) 2HDM: 
left(right).
}
\label{fig2}
\end{figure}
Measurements at LHC, ILC and PLC are complementary, being sensitive to 
different  Higgs-boson couplings    \cite{Niezurawski:2004ui,Weiglein:2004hn}. 
This is shown in Fig. 3, left,  where the ratios of the  cross sections times 
BR to WW, 
obtained in the CP conserving 2HDM and SM, are presented
as functions of $\chi_V^H$ and $\chi_u^H$ for $M_H$ = 250 GeV 
($M_h$ = 120 GeV, $M_{H^+}$ = 800 GeV). 
Results for 2HDM(II) with weak CP violation via  H-A mixing 
 show  similar compementarity (Fig.3, right). Here measurements of 
 mixing phase $\phi_{HA}$ and the $\chi_u^{(2)}$ for LHC, and $\chi_V^{(2)}$  
 at ILC and PLC, are presented for $h_2$  with mass 250 GeV.

Simultaneous fit to the   invariant
mass distribution for $W^+W^-/ZZ$ at  LHC, ILC and PC was performed 
with 12 parameters: 
$\chi_ V,\chi_u, M_H,\phi_{HA}$ and 8 parameters describing normalization and
PLC luminosity spectra shape, corresponding to following
systematic uncertainties:
signal normalization at LHC (ILC,PLC): 20\%(5,5),
 background normalization at LHC (ILC,PLC): 10\%(5,10), and
   PLC spectra shape:  10\%.

In general case the combined analysis of LHC, ILC and PLC data is needed.

\vskip 1cm
{\bf{Acknowledgment:}} I am grateful to Ilya Ginzburg, David Temes,
A. Filip  \.Zarnecki and  Piotr Nie\.zurawski  for fruitful collaborations 
and contributions to this summary.
{Supported in part
  by the Polish Committee for Scientific Research,
   grant  no.~1~P03B~040~26 
   and project no.~115/E-343/SPB/DESY/P-03/DWM517/2003-2005.}


\begin{thebibliography}{99}
\bibitem{Ginzburg:2004vp}
  I.~F.~Ginzburg and M.~Krawczyk,
  arXiv:hep-ph/0408011, Phys. Rev. D to appear
\bibitem{Krawczyk:2004na}
  M.~Krawczyk and D.~Temes,
  Eur.\ Phys.\ J.\ C {\bf 44}, 435 (2005)
  [arXiv:hep-ph/0410248].
\bibitem{Gambino:2001ew}
  P.~Gambino and M.~Misiak,
  Nucl.\ Phys.\ B {\bf 611}, 338 (2001)
  [arXiv:hep-ph/0104034].
\bibitem{Niezurawski:2005cr}
  P.~Niezurawski, A.~F.~Zarnecki and M.~Krawczyk,
  arXiv:hep-ph/0507006.
\bibitem{Niezurawski:2004ui}
  P.~Niezurawski, A.~F.~Zarnecki and M.~Krawczyk,
  JHEP {\bf 0502}, 041 (2005)
  [arXiv:hep-ph/0403138],
  Acta Phys.\ Polon.\ B {\bf 36}, 833 (2005)
  [arXiv:hep-ph/0410291],
LC-PHSM-2004-035
\bibitem{Weiglein:2004hn}
  G.~Weiglein {\it et al.}  [LHC/LC Study Group],
  arXiv:hep-ph/0410364.
  \end{thebibliography}
\end{document}